\documentclass[fleqn,usenatbib]{mnras}

\usepackage{newtxtext,newtxmath}
\usepackage[T1]{fontenc}

\DeclareRobustCommand{\VAN}[3]{#2}
\let\VANthebibliography\thebibliography
\def\thebibliography{\DeclareRobustCommand{\VAN}[3]{##3}\VANthebibliography}

\usepackage{graphicx}	% Including figure files
\usepackage{amsmath}	% Advanced maths commands
\newcommand{\bicho}{4FGL J1848.7--0129}

   \title{A blazar candidate for the Fermi source \bicho}

 \author[P.L. Luque-Escamilla et al.]{
   Pedro L. Luque-Escamilla,$^{1}$\thanks{E-mail: peter@ujaen.es (PLLE)}
   Josep Mart{\'i},$^{2}$
   Enrique  Mestre,$^{2}$
   Jorge A. Combi,$^{2,3,4}$ 
   Juan F. Albacete-Colombo,$^{5}$
\\
$^{1}$ Departamento de Ingenier{\'i}a Mec{\'a}nica y Minera (EPSJ), Universidad de Ja{\'e}n, Campus Las Lagunillas s/n, A3, E-23071 Ja{\'e}n, Spain\\
$^{2}$Departamento de F{\'i}sica (EPSJ), Universidad de Ja{\'e}n, Campus Las Lagunillas s/n, A3, E-23071 Ja{\'e}n, Spain\\
$^{3}$Instituto Argentino de Radioastronom\'ia (CONICET; CICPBA), C.C. No 5, 1894 Villa Elisa, Argentina\\
$^{4}$Facultad de Ciencias Astron\'omicas y Geof\'isicas, Universidad Nacional de La Plata, Paseo del Bosque, B1900FWA La Plata, Argentina\\
$^{5}$Universidad de R\'io Negro, Sede Atl\'antica - CONICET, Viedma CP8500, R\'io Negro, Argentina
}

\date{Accepted XXX. Received YYY; in original form ZZZ}

\pubyear{2015}

\begin{document}
\label{firstpage}
\pagerange{\pageref{firstpage}--\pageref{lastpage}}
\maketitle

\begin{abstract}

The Fermi source \bicho\ has been historically related to the globular cluster GLIMPSE-C01 since its very first detection.
Although this association is widely accepted, as it appears in the most recent Fermi catalog, it deserves to be revisited given the multi-wavelength evidences and the recent discovery of variable X-ray sources in the Fermi source region. In particular, low frequency radio maps from the Giant Metre Radio Telescope in Pune (India) have been carefully inspected which, together with X-ray data re-analysis from Chandra, lead us to get a deep insight into the candidates to be associated to \bicho. This results in the discovery of a new X-ray variable point source coincident with an unreported non-thermal radio emitter, both of them well inside the \bicho\ error ellipse. We analyze and discuss all these observational facts, and we propose now a newly discovered blazar candidate as the most promising responsible for the gamma ray emission in the Fermi source. If confirmed, this result would set constrains on the number of millisecond pulsars in GLIMPSE-C01 or their gamma-ray emission properties.
\end{abstract}


%%%%%%%%%%%%%%%%%%%%%%%%%%%%%%%%%%%%%%%%%%%%%%%%%%

%%%%%%%%%%%%%%%%% BODY OF PAPER %%%%%%%%%%%%%%%%%%

\section{Introduction}

More than a decade ago, the collaboration operating the Fermi Large Area Telescope (LAT) released their first catalog based only on three months of gamma-ray observations \citep{0FGL}. One of the sources was listed as 0FGL J1848.6$-$0138, whose nature remained unknown. In spite of its large 95\% confidence error radius of $\sim 10$ arcmin, some of the present authors were the first to notice \citep{2009LuqueG01} that the gamma-ray emission coming from this region of the sky could be attributed to the highly absorbed globular cluster GLIMPSE-C01  \citep[hereafter G01;][]{2005Kobulnicky}, located at a distance of few kpc, ranging from $4.2$ kpc \citep{2016Davidge} to $5.0$ kpc, \citep{2011Davies}. Its mass has been estimated to be $2.81 \times 10^4$ M$_\odot$, while its age seems to lie between 0.3 and 2 Gyr \citep{2011Davies, 2018Hare}, which could suggest that G01 is an intermediate-age stellar cluster instead of a typical, older globular one. Nowadays, the same gamma-ray source appears in the more recent Fermi LAT catalog \citep{2022_4FGLDR3} as \bicho\, with a much smaller ellipse error of $\sim 2$ arcmin after more than 12 years of observations. The globular cluster association has received increased reliability, with the G01 position being well coincident with the gamma-ray emission, and so it appears in the catalog. The high-energy photons coming from G01 could be collectively produced by a population of millisecond radio pulsars (MSPs) inside the cluster through injection of relativistic leptons into the medium either from their inner magnetospheres or accelerated in the shocks created by the collision of individual pulsar winds \citep{2009LuqueG01, 1993tavani}. Moreover, G01 has been used to study the gamma-ray MSP population of globular clusters in order to explain the GeV excess observed from the region surrounding the Galactic Center  by \cite{2016Hooper} and, more recently, by \cite{2022Wu}, who estimate 1-6 MSPs inside G01. 
%(with 3 giving the best PL cutoff fit) según Wu et al, with minimum chisquared of 5.3 for 5 data points (limit of 5\% smallest values of chisquared is chisquared = 6.3)

However, in the recent past there has been much activity in the astrophysical community related to G01. On December 2020 the Monitor of All-sky X-ray Image (MAXI) discovered the new X-ray transient MAXI J1848--015 \citep{2020ATel14282}, which was subsequently detected by NuSTAR \citep{2020ATel14290,2022Pike} and, with an improved position, by Swift \citep{2021ATel14420}, NICER \citep{2021ATel14429}, MeerKAT \citep{2021ATel14432} and Chandra \citep{2021ATel14499, 2021ATel14424}. Thus, MAXI J1848--015 seems to be located in the core of G01, with no coincidence with known archival sources. 

All these findings prompted us to revisit the gamma-ray association between \bicho\ and G01, taking into account all the multi-wavelength data available up to now. As a result, here we propose a new likely non-thermal radio emitter, coincident with the Fermi error ellipse and a X-ray Chandra point source, that could be contributing to the gamma-ray emission.
%to alternatively be the responsible for the gamma-ray emission instead of MSPs in G01. 
As illustrated in the following sections, this previously unreported object seems to be a member of the so-called blazars, a radio-loud subclass of active galactic nuclei that host relativistic jets closely aligned with our line of sight \citep{1995Urry}. They are characterized by strong non-thermal radiation across the entire electromagnetic spectrum, and in particular in gamma-rays, with fluxes expected to be between $6\times 10^{-10}$ and $2\times 10^{-6}$ photons cm$^{-2}$ sec$^{-1}$ in the $0.1-100$ GeV range in Fermi blazars \citep[i.e.][]{2015singal}. Blazars also exhibit strong variability at different timescales and wavebands, believed to be a result of relativistic motion of non-thermal plasma along the jet, which dominates the blazar emission due to relativistic beaming because of its particular orientation \citep{1995Urry}. The multiwavelength spectral energy distribution (SED) of a blazar exhibits a typical double hump structure. The low-energy bump peaks at radio-to-X-rays and is believed to be formed by synchrotron mechanism. On the other hand, high-energy bump peaks in the gamma-ray band and is usually explained in terms of inverse Compton process \citep[in a leptonic radiative model,][]{2019vdB} or hadronic processes \citep[\textit{e.g.},][]{2015petro}.

% GC01 is of interest for studies of the evolution of the Galactic disk because it is one of the most massive clusters that may have formed during intermediate epochs (Davies et al. 2011). The formation of large, compact clusters is often associated with interactions and/or starburst events (e.g., Ashman \& Zepf 2001). Does the age of GC01 coincide with a past event that may have influenced Galactic evolution? Davies et al. (2011) conclude that GC01 is not an old globular cluster, and assign it an age between 0.3 and 2 Gyr, with the most probable age between 0.4 and 0.8 Gyr. Davies et al. (2011) further suggest that the formation of GC01 may be linked to a past encounter between the Galactic disk and the Magellanic Clouds. In fact, Rezaei et al. (2014) find peaks in the SFRs of the LMC and SMC $\sim 0.7$ Gyr in the past, which they suggest may be linked to an interaction with the Galaxy. However, other studies suggest that GC01 may be too old to have formed as part of the most recent interaction with the Magellanic Clouds, and claim that the presence of young clusters with masses $>10^4$ Msun like Westerlund 1 and the Arches suggests that clusters with masses approaching that of GC01 may form naturally throughout the lifetime of the Galaxy, without the need for an external trigger (Davidge et al 2016).  GLIMPSE C01 does not have number of stellar encounter rates available. 

\section{Observations}

We have reanalysed archival observations at different waveleghts (see Table\ref{tab:obs}). Next, we show the corresponding results.

\begin{table}
\caption{Observations data log.}
\label{tab:obs}
\begin{tabular}{lccc}
\hline
    & Observatory / Inst. & ObsID. & Date\\
%& $M_{\sun}$ & $L_{\sun}$\\
\hline
Radio (150 MHz) & GMRT & TGSS &15/03/2016\\
Radio (333 MHz) & MAGPIS &  & 02/09/2001\\
Radio (1400 MHz) & MAGPIS & & 25/04/2007\\
X-ray (0.5-7.0 keV) & Chandra ACIS-S & 6587 & 15/08/2006\\
 & Chandra / ACIS-I & 21641 & 23/06/2019\\
 & Chandra / ACIS-I & 21642 & 28/06/2020\\
 & Chandra / ACIS-I & 21643 & 14/07/2020\\
 & Chandra / ACIS-I & 21644 & 08/08/2020\\
 & Chandra / ACIS-I & 21645 & 05/08/2020\\
$\gamma$-ray (0.1-500 GeV) & Fermi - LAT &   & 04/08/2008 to\\
                               &             &   & 28/03/2022\\
\hline
\end{tabular}
\end{table}

\subsection{Radio data}

\begin{figure}
\includegraphics[width=8.0cm, angle=0]{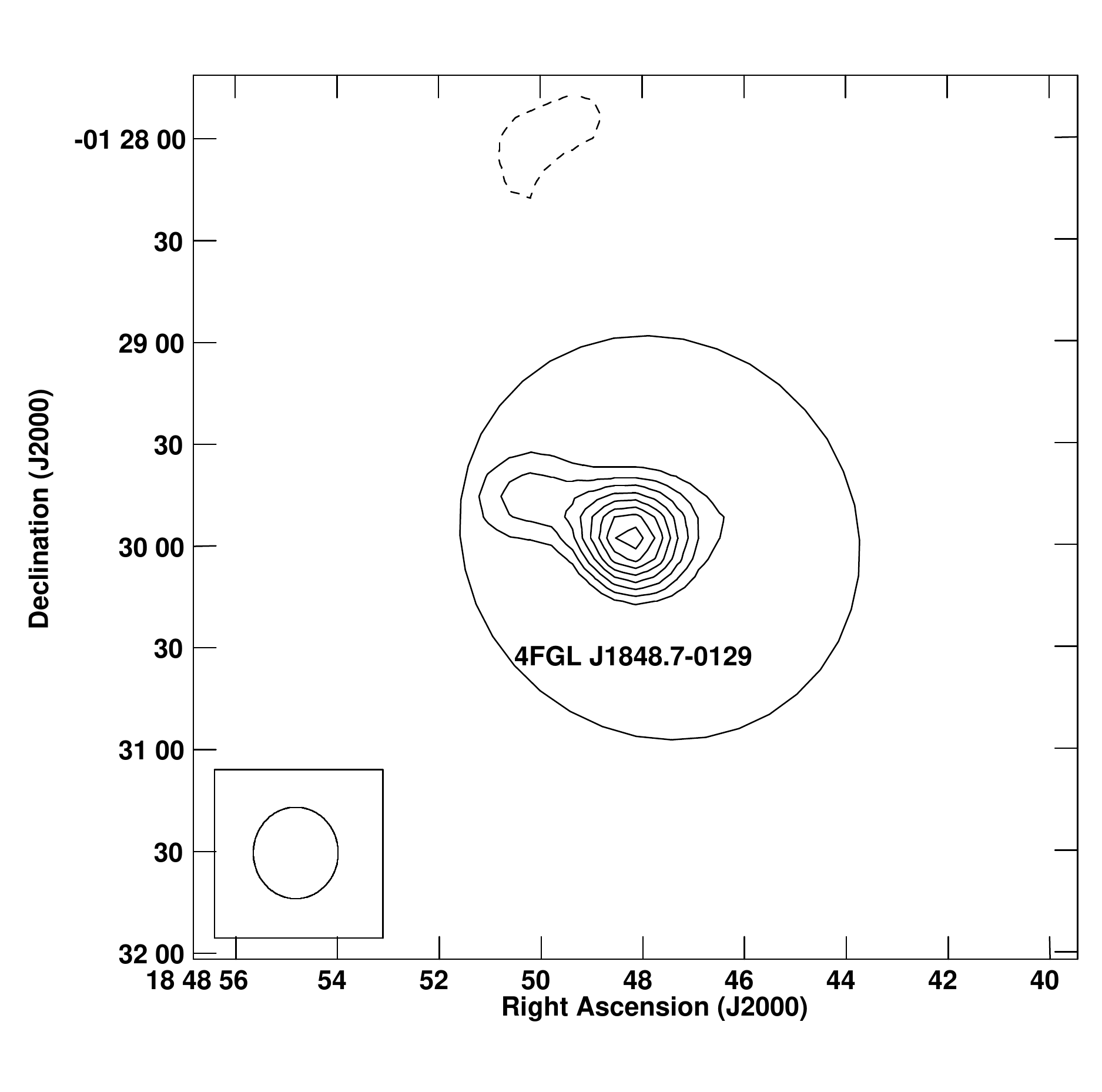}
%\imageii
\caption{   \label{GMRT}  Radio emission present in the GMRT 150 MHz All-sky Radio Survey towards the
gamma-ray source  4FGL J1848.7-0129 whose 95\% confidence ellipse is also plotted.
The contours shown correspond to -3, 3, 4, 5, 6, 7, 8, 9 and 10 times 7 mJy beam$^{-1}$, the rms noise.
The small bottom left ellipse illustrates the GMRT clean restoring beam of $27\times 25$ arcsec$^2$, with position angle of $0^{\circ}$.
}%% no full stop at the end
\end{figure}

\cite{2005Kobulnicky} first analyzed the NRAO VLA Sky Survey \citep[NVSS,][]{1998AJ....115.1693C} radio data coming from G01 region. A marginal and extended detection with an integrated flux density of $20.5 \pm 3.6$ mJy at 20 cm was coincident with the cluster. The present authors recalibrated the Very Large Array (VLA) archive 20 cm data available in B array configuration in order to obtain a higher angular resolution radio map. As a result, no compact radio sources were detected above four times the root mean-square noise of $0.25$ mJy beam$^{-1}$, thus concluding that radio emission was intrinsically extended or produced by the combined effect of faint point-like radio sources \citep{2009LuqueG01}.
%Radio emission towards the globular cluster GLIMPSE C01 has been pointed out in the past at a faint and diffuse level, as in  \citet{2009LuqueG01} based on images from theNRAO VLA Sky Survey (NVSS, \cite{1998AJ....115.1693C}). 
The Galactic Plane radio emission at so low latitudes often renders  difficult to disentangle weak and confused radio sources.
This circumstance prompted us to inspect other newer interferometric radio surveys that also cover G01 field of view but with improved angular 
resolution. An unexpected finding appeared when using the
First Alternative Data Release TGSS ADR1 of the
Giant Metre Radio Telescope (GMRT) 150 MHz All-sky Radio Survey \citep{2017A&A...598A..78I}. Here, a conspicuous low-frequency radio source 
 stands out inside the 95 \% confidence ellipse of the gamma-ray source 4FGL J1848.7-0129  currently associated with G01 (see Fig. \ref{GMRT}).
 %The AIPS tasks MAXFIT and JMFIT were used to measure its main observed parameters.
 The J2000.0 coordinates of the radio peak
 position correspond to RA = 18:48:48.23 and DEC = -1:29:57.6, with an estimated uncertainty of 2.6 arc-second in each coordinate.
 This location is offset from the cluster core by almost half arc-minute, thus not being coincident with
 the recently reported MeerKAT point-like radio source associated to MAXI J1848-013  \citep{2021ATel14432}.
 The GMRT 150 MHz peak flux density is $74 \pm 7$ mJy while the total flux density rises to $110 \pm 24$ mJy, thus showing that our new radio
 source is clearly extended. This is also seen in Fig. \ref{GMRT} where the GMRT object appears elongated roughly in the East-West direction.
 The fact that this source was missed in the associated GMRT survey catalog is probably due to its signal-to-noise ratio (${\rm SNR} \sim 10$) 
 being not far from the conservative threshold chosen by its authors.
 
 Another improved radio view of the G01 region is provided by the 
 Multi-Array Galactic Plane Imaging Survey (MAGPIS) as described in \citet{2006AJ....131.2525H}. Hints of emission at the $\sim 0.1$ Jy level are present in the MAGPIS maps at the 90 cm wavelength close to the GMRT radio source position, while at the 20 cm wavelength no obvious MAGPIS detection is present at the GMRT source position with a $4\sigma$ upper limit of 5.8 mJy. However, a weak emission at 20 cm in MAGPIS seems to be coincident with the East-West elongated feature in GMRT map. This limb of radio emission might be associated with MAXI J1848-015 or diffuse radio emission from the cluster.

\subsection{X-ray data}

\cite{2007Pooley} studied Chandra data from G01 using a single 46 ksec  ACIS observation. They report 17  point-like sources inside the 36 arcmin radius half-light of the cluster, which are related to a mixture of cataclysmic variables, quiescent low-mass X-ray binaries (LMXB), and MSPs, among other objects. 
%Remarkably, the source (\#X17) in that survey is spatially coincident with the peak of the radio source here reported. 

To obtain a more precise detection and analysis of this X-ray source, we use a complete set of 7 ACIS observations (ObsId. 6587, 21641, 21642, 21643, 21644, 21645, and 21646) that leads to a total of exposure 288.87 ksec, more than six times deeper than \cite{2007Pooley} study.
The last provided CIAO 4.14 version and the CALDB 4.9.6 set of calibration files were used. When applying the CIAO \textit{wavdetect} task for source detection in the entire 0.5–8.0 keV broadband, we detected 30 X-ray point sources on a restricted field of view of 3.9$\times$3.5 arcmin$^2$ side centered at the G01 position.
%(RA = 18:48:48, DEC = -01:29.47). 

%\begin{figure}
%\includegraphics[angle=0,width=9.5cm, angle=0]{specX6_mean.png}
%Note: Non-thermal fitted model (blue) to the X6 observed X-ray spectrum. Red models corresponds to the 1$\sigma$ bottom and upper limit  solutions according to N$_{\rm H}$ and $\Gamma$-index errors. Adjusted model converge for log(N$_{\rm H}$)=22.9($\pm$0.3) cm$^{-2}$, $\Gamma$=2.2($\pm$0.6), z=0.3 (fixed) and normalization of 2.1$\times$10$^{-6}$ cm$^{-5}$.
%\label{x:spectrum}
%\end{figure}

\begin{figure}
\centering
\includegraphics[width = \linewidth]{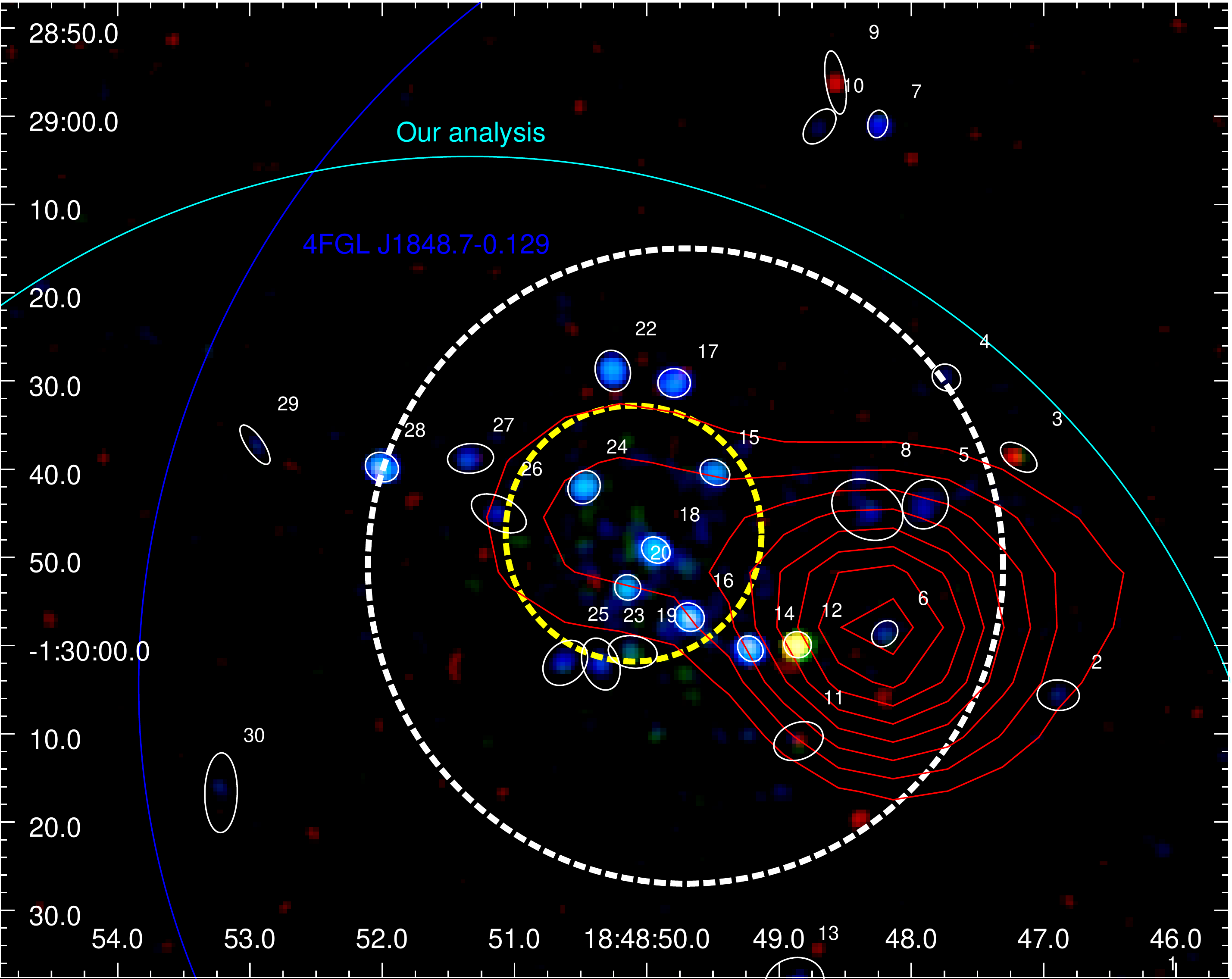}
\caption{The full X-ray band is coded in color so that soft [0.5 – 1.2] keV emission appears in red, and medium [1.2 – 2.5] keV in green, while hard [2.5 – 8.0] keV emission appears in blue. The GLIMPSE-C01 lie inside the 4FGL $\gamma$-ray source ellipses (blue) and our own analysis of the Fermi observation (cyan). Red contours represent the GMRT radio data at 150 MHz.  White and yellow dashed-circle means the 36" and 14.5" effective radius of the cluster in X-rays \citep{2007Pooley} and in the infrared \citep{2011Davies}, respectively.}
\label{fig:Chandra}
\end{figure}

One of these X-ray sources deserves special attention. It is the one labeled as X6 in our survey (see Fig. \ref{fig:Chandra}), which is remarkably coincident with source X17 in Pooley et al paper. Its location is RA = 18:48:48.20 and DEC = -1:29:58.7 J2000.0, with an uncertainty of 0.3 arc-second in each coordinate. Therefore, it is consistent with the peak of the GMRT radio emission (see Fig. \ref{fig:Chandra}), while it is not coincident with MAXI J1848-015. In fact, in this figure we can see that our X18 source is the same than that associated with MAXI J1848-015 in Chandra  \citep{2021ATel14424} and MeerKAT \citep{2021ATel14432}. X6 has 16 X-ray photons from the stacked observations, which imposes a statistical limit for an X-ray spectral fitting. However, for our analysis, we will take advantage of the results of \cite{2016BAlbacete}, who use Chandra Acis Montecarlo Photon Estimator Recipe (Camper)\footnote{ess: http://camper.lia.unrn.edu.ar} routine to perform a statistical assessment of the X-ray spectral fitting procedure.
%\citep{2016Albacete}. 
We simulated a 16 photons spectrum and used a non-thermal absorbed model (TBABS $\times$ PO) from \citet{2011Arnaud}. If the source were located in the Galaxy, we can take as a lower limit for the neutral absorption column of Hydrogen the value from \citet{2007Pooley} N$_{\rm H}$ = 2.7$\times$10$^{22}$ cm$^{-2}$. On the other hand, an upper limit for this parameter could be obtained if we take into account that the source is a blazar without optical counterparts (see Section 2.4) but detected in X-rays, 
%as it would be better characterized by 
where N$_{\rm H}$ < 1.5$\times$10$^{24}$ cm$^{-2}$ for redshifts ($z$) < 0.5 \citep{2012Mateos}. Therefore, a value 4 or 5 times that adopted in \citet{2007Pooley}, such as N$_{\rm H}$ = 10$^{23}$ cm$^{-2}$, could be a reasonable assumption for our object. To fine-tune this, we can use an spectral model. As no infrared counterpart is observed (see Section 2.4), we assume the blazar to be a Flat-Spectrum Radio Quasar (FSRQ) source with X-ray emission well described by a non-thermal spectrum dominated by jet synchrotron emission \citep{2008Landt}. So, we assume a power-law emission model of index $\Gamma \simeq 2.0$ to simulate the X-ray spectrum with 16 photons. As a result, we get log(N$_{\rm H}$)=22.9($\pm$0.3), $\Gamma$=2.2($\pm$0.5), with a normalization of 2.1$\times$10$^{-6}$ in cm$^{-5}$. The corresponding corrected absorbed flux for the 2.0 to 10.0 keV energy range is log(Flux) = -13.98 ($\pm$ 0.15) erg\,s$^{-1}$\,cm$^{-2}$.

%The source is certainly towards G01, where the neutral absorption column of Hydrogen is N$_{\rm H}$ = 2.7$\times$10$^{22}$ cm$^{-2}$ according to \citet{2007Pooley}. This may be taken as a lower limit constraint for the N$_{\rm H}$ of the source. Most AGNs without optical counterparts but detected in X-rays, like this source, are well characterized by N$_{\rm H}$ under 1.5$\times$10$^{24}$ cm$^{-2}$ for redshifts ($z$) under 0.5 \citep{2012Mateos}. The assumption of an N$_{\rm H}$ value over 4 or 5 times that adopted in \citet{2007Pooley} (N$_{\rm H}$ = 10$^{23}$ cm$^{-2}$) would be a plausible assumption for our object. Otherwise, a power-law emission model of index $\Gamma$ of 2.0 is well representative of High-state Blazars (HBLs), as their emission mostly occurs in the UV/X and $\gamma$-ray bands, as is observed \citep{2021Middei}. Under these restrictions, and for a simulated X-ray spectrum with 16 photons, we get log(N$_{\rm H}$)=22.9($\pm$0.3), $\Gamma$=2.2($\pm$0.5), normalization of 2.1$\times$10$^{-6}$ in cm$^{-5}$. So, the corrected absorbed flux log(Flux) = -13.98 ($\pm$ 0.15) erg\,s$^{-1}$\,cm$^{-2}$ is obtained for the 2.0 to 10.0 keV energy range.

%We also search for variability of source X6 using a classical Kolmogorov-Smirnov (KS) test to the X-ray lightcurve with 30 ksec binning (Fig. \ref{fig:LATlc}). As a result, a p-value $p_{KS, X} \simeq 0.01$ was obtained, meaning that the data taken has a probability of $\sim 99\%$ of coming from a variable source. 

\noindent

In order to test the variability of the source we have built a light-curve based on the precise count-rate of \textit{Chandra} (see Fig. \ref{fig:LATlc}).
Finally, we have also sought for more X-ray data in different observatories. We found that NuSTAR has detected \citep{2020ATel14290} a source which has been related to the X-ray transient MAXI J1848--015 \citep{2020ATel14282,2022Pike}. This NuSTAR emitter is far from our Chandra X6 source and will not be taken into account in our discussion.

%\textbf{ Swift?? Debe haber datos relacionados con la MAXI. Basta ver que no son coincidentes con la emisi\'on radio para desecharlos, pero debemos comprobarlo.}

\subsection{Gamma-ray data}

We also revisited the LAT data towards 4FGL J1848.7-0129 for a region of $15\degr{}$ radius centered at RA = 18:48:48.0 and DEC = -1:30:00.0. We employed all LAT {\sc SOURCE}-quality events recorded after more than 13 years of data taking (from August 4, 2008, to March 28, 2022) at energies from 100 MeV to 500 GeV with 90$\degr{}$ of maximum zenith angle (to prevent contamination from Earth limb events). We then analyzed the LAT data applying the joint likelihood fitting method provided by the \textsc{fermipy} \textsc{python} package  \citep[version 1.0.1, built upon the \textsc{Fermi Science Tools};][]{2017ICRC...35..824W}.
The model fitted to data includes all the LAT sources listed in the Fermi-LAT Fourth Source Catalog \citep[4FGL;][]{Abdollahi_2020} in $20\degr{}$ radius around the reference position above, plus the Galactic and extra-galactic diffuse gamma-ray components described with the most recent version of the Galactic (gll\_iem\_v07) and isotropic (iso\_P8R3\_SOURCE\_V3\_v1) diffuse emission models. We evaluated the LAT instrument's response with the version P8R3\_SOURCE\_V2 of the instrument response functions, applying the energy dispersion correction to all sources except for the isotropic diffuse emission. The model's free parameters are; (1) the normalization parameter of all sources with a detection significance above 3$\sigma$ ($\sqrt{\rm TS} > 3$), (2) all spectral parameters regarding the sources in $5\degr{}$ radius around the source's reference position, and (3) the isotropic (norm) and Galactic (norm and tilt) diffuse emission parameters. Lastly, we estimated the systematic uncertainties due to the LAT effective area and the diffuse Galactic model.

The LAT source 4FGL J1848.7-0129 was located (as point-like), with large detection significance ($\sqrt{\rm TS} \approx 30$), at the position RA = 18:48:51.36 $\pm$ 28.8$^{\prime\prime}_{\rm stat}$ $\pm$ 30.3$^{\prime\prime}_{\rm sys}$ and DEC = -1:30:36.0 $\pm$ 32.4$^{\prime\prime}_{\rm stat}$ $\pm$ 21.1$^{\prime\prime}_{\rm sys}$. We compared the best-fit Gaussian and radial disk source hypotheses with the point-like assumption through the likelihood ratio test, resulting in no significant evidence of extended emission. The spectral energy distribution (SED) of the source was modeled with a log-parabola function ($dN/dE = N_{0} \times (E/E_{\rm ref})^{-(\alpha +\beta \log (E/E_{\rm ref}))}$), commonly used to characterize blazar spectra at high-energy gamma-rays, for 13 energy bins spanning from 100 MeV to 500 GeV. The model's parameters that best represent data consist of; $N_{0} = (8.73 \pm 0.30_{\rm stat} \pm 0.45_{\rm sys}) \times 10^{-13}$ cm$^{-2}$ s$^{-1}$ MeV$^{-1}$ referenced at $2.91$ GeV, $\alpha = 2.53 \pm 0.03_{\rm stat} \pm 0.06_{\rm sys}$, and $\beta = 0.33 \pm 0.02_{\rm stat} \pm 0.08_{\rm sys}$. 
% $\rm{RA} = 282.214\degr{} \pm 0.008\degr{}$ and $\rm{DEC} =  -1.510\degr{} \pm 0.009\degr{}$
   
\cite{2016Hooper} and \cite{2022Wu} studied the LAT data towards G01 to quantify the high-energy contribution of its MSPs, assuming that the gamma-ray emission exhibits an exponentially cutoff power-law spectral shape (typical of MSPs). \cite{2021Song} found that globular clusters spectra can be better resolved into an exponentially cutoff power-law together with a pure power-law model, which is most naturally interpreted as inverse Compton emission by cosmic-ray electrons and positrons injected by MSPs. To probe this scenario, we fitted an exponentially cutoff power-law model to the LAT data examined, resulting in cutoff energy of $4.19 \pm 0.15_{\rm stat} \pm 0.34_{\rm sys}$ GeV. However, it does not improve the log-parabola fit detailed above (at energies between 100 MeV and 10 GeV), according to the Akaike's Criterion \citep[as we get AIC$_{\rm LP}$ - AIC$_{\rm ECPL}=-6.5 < 0 $;][]{Akaike1973InformationTA,1100705}.

\begin{figure}
\centering
\includegraphics[angle=0,width = \linewidth]{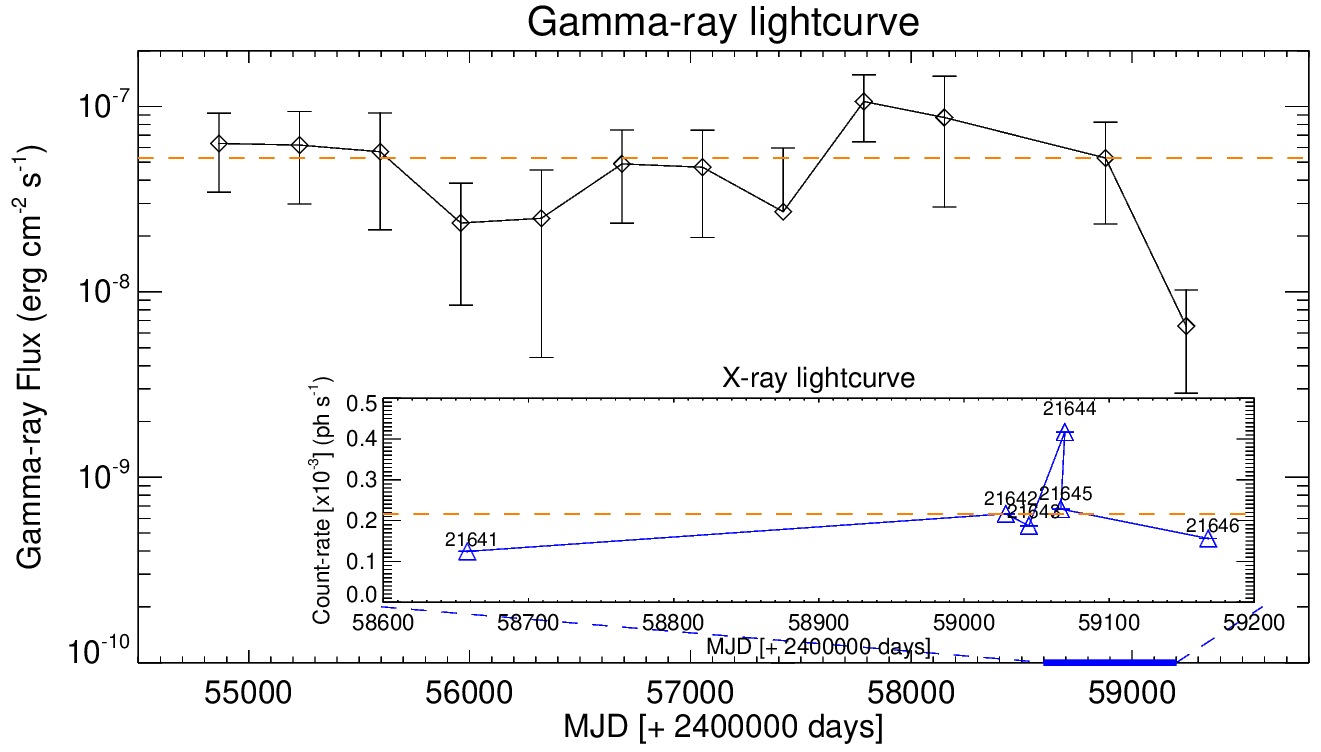}
\caption{Left: Long-term 100 MeV to 500 GeV $\gamma$-ray light curve of 4FGL J1848.7-0129 as extracted from the 4FGL second release. The time bin-size is of one year long. The inner-left panel shows the X-ray light curve at a shorter time scale. Both line-dotted lines indicate the median values of 5.28$\times$10$^{-8}$ erg cm$^{-2}$ s$^{-1}$ and 0.21$\times$10$^{-3}$ photons s$^{-1}$ for the $\gamma$ and X-ray, respectively. Note: Labels over X-ray symbols refers to the observation Id. numbers of the Chandra data. Error bars are smaller than the symbol. }
\label{fig:LATlc}
\end{figure}

The variability of the LAT source was studied in timescales ranging from four months to three years by computing its light curve with time bins of various sizes (see Fig. \ref{fig:LATlc}). The smallest binning used is that in which the average source's detection significance for the different time bins is $\sim 5\sigma$.

%\begin{figure}
%\centering
%\includegraphics[angle=0,width=9cm, angle=0]{4fgl_j18487_0129_lightcurve.png}
%\caption{The light curve of 4FGL J1848.7-0129 at energies from 100 MeV to 500 GeV with time bins of four months of size.}
%\label{fig:LATlc}
%\end{figure}

%gamma flux of $9.020 \times 10^{-12}$ (but with uncertainty of $13\%$ (+1.205, -1.345), Hoope y Linden 2016) following a model $dN_\gamma / dE_\gamma \propto E^{-\alpha} \exp(-E_\gamma/E_{cut})$, with $\alpha = -0.74$ and $E_{cut} = 1.58$ GeV with test statistic TS = 68.7 (Hoope y Linden 2016).

\begin{figure}
\centering
\includegraphics[angle=0,width=9cm, angle=0]{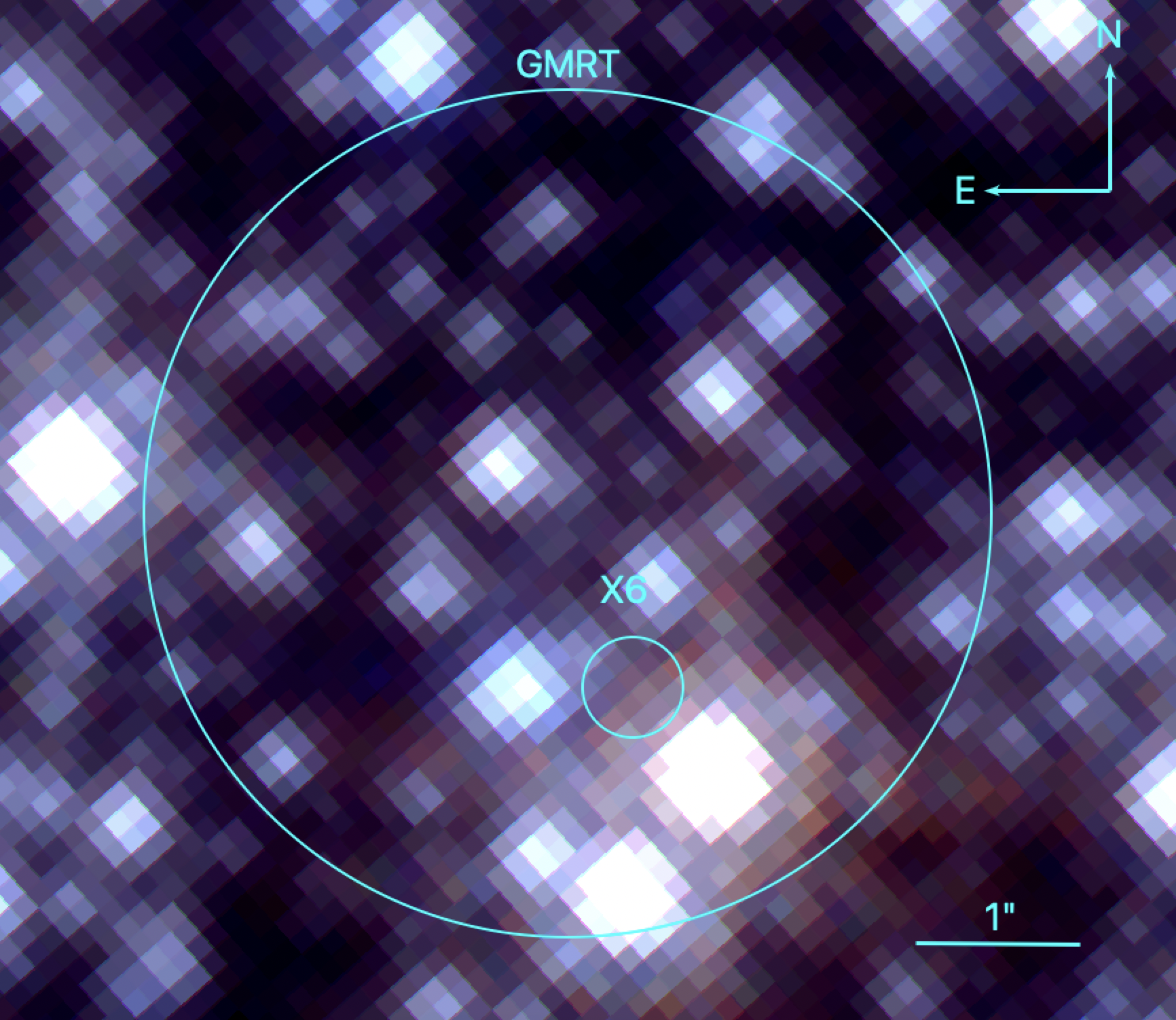}
\caption{Tri-chromatic view of G01 GMRT radio peak position, and Chandra X-rays X6 source position, as observed with the Hubble Space Telescope WFC3. The red, green and blue layers correspond to the F153M, F139M and F127 M filters, respectively.}
\label{fig:HST}
\end{figure}

\subsection{Archival optical/infrared data}
Although we expect a very intense extinction, we seeked for optical/infrared counterparts of our X6 X-ray point source.  We retrieved the archival Hubble data obtained with the Wide Field Camera 3 (WFC3) previously reported by \citep{2018Hare} in quest for possible counterparts for further supporting the later discussion. Astrometry has been linked to the Gaia EDR3 reference stars. These Hubble product also come with photometric calibration available in the header of the fits files. At the Hubble WFC3 wavelengths (12.7, 13.8 and 15.3 $\mu$), the corrections for Galactic interstellar absorption assuming the \citet{2007Pooley} $N_H$ value is almost negligible ($\sim 0.02$ magnitudes or less). As a result, no infrared counterparts are apparent in the position of the X6 source (see Fig \ref{fig:HST}). The same is true when inspecting the optical bands of the $2^{nd}$ Digitized Sky Survey (DSS2) catalogue.

\section{Discussion}
We first stress the coincidence between the X-ray source X6 and the peak of GMRT radio emission. The offset between the respective coordinates is within the astrometric error at the arc-second level. Therefore we proceed under the assumption that both sources are the same object. No infrared counterpart is detected (see Fig. \ref{fig:HST}), and this is likely due to the strong absorption towards this direction as suggested by the X-ray measurements. 

In order to constraint the radio spectral index, we assume constant flux densities for the newly reported source, and a power law spectrum (with flux $f_\nu \propto \nu^{-\alpha}$). As a result, from the MAGPIS 90 cm and GMRT maps we roughly estimate an almost flat spectral index $\alpha \simeq 0.0$, which is in accordance to the typical average value at radio band emission in blazars \citep{Abdo2010_SED}. If we take into account the lack of detection in MAGPIS 20 cm maps, we get a significantly high $\alpha \geq 1.1$, which suggests that the constant flux density assumption is possibly not applying, and the source is variable in radio and lying in a low state at the time of the MAGPIS 20 cm observation.

This hint of variability is reinforced if we analyze the high-energy light curves of our candidate. We will use a common one-sample Kolmogorov-Smirnov (KS) test to compare the observed data with a uniform distribution, which does not depend neither on the binning nor on the size of the sample \citep[see, \textit{i.e.}, ][]{1990Daniel:ch8.2}. According to this test, in the X-ray domain  we get a p-value $p_{KS,X} \simeq 0.013$, which means that the source is not constant with a $98.7\%$ probability. This value is high enough to assure the transient nature of the observed X-ray flux with statistical significance, albeit the scarcity of the sample, The X-ray variability seems to be visually confirmed in the lightcurve in Figure \ref{fig:LATlc}, although we take it with due caution.  In the gamma-ray lightcurve the KS test results in a p-value $p_{KS,\gamma} \simeq 0.31$, which cannot let us to reject the null hypothesis of a constant source with a high significance. However, this implies that the $70\%$ probability of the observed fluxes come from a variable emitter, which is not negligible. Therefore, in order to obtain an independent evidence of variability for LAT data, for each gamma-ray lightcurve we computed the \textit{normalized excess variance} quantity \citep{2003MNRAS.345.1271V} that is commonly used to estimate the variability amplitude in AGN gamma-ray light curves \citep[see, e.g.,][]{2010ApJ...722..520A, 2020A&A...634A..80R}. It is defined as $\sigma_{NXS}^{2} = (S^{2} - \langle\sigma_{err}^{2}\rangle)/\langle F \rangle^{2}$, where $S^{2}$ stands for the light curve's variance, $\langle F \rangle$ for its average, and $\sigma_{err}^{2} = \sigma_{\rm stat}^{2} + \sigma_{\rm sys}^{2}$ are the systematical and statistical errors added in quadrature (we used $\sigma_{\rm sys} = 0.03\times \langle F \rangle$). However, we obtained near zero, but not positive values for $\sigma_{NXS}^{2}$ for all the cases, indicating either very little variability in monthly or yearly timescales or slightly overestimated errors. The latter is likely our case (as apparent in Fig. \ref{fig:LATlc}), given the large errors obtained with $\sigma_{i}/F_{i} \gtrsim 0.4$ in most bins.  In any case, this lack of remarkable variability in blazars in gamma-rays is not surprising. The Fermi 4FGL DR3 catalog \citep{2022_4FGLDR3} contains 1410 sources labeled as 'bll', $i.e$, cataloged as blazars, from which only 504 (36\%) have a variability index with significance over 99\%. From the remaining 906 sources, there is not a well-defined fractional variability index in 272 (26\%), so we can conclude that a neat variability is not always evidenced at these energies.

Although with due caution, we  suspect that we are dealing with a highly obscured blazar source.
%, given the above hints of a non-thermal and variable source in the both X-ray and gamma-ray domains. 
This is relevant because it opens a new perspective towards the identification of \bicho\ that could be even completely unrelated to G01 as is widely accepted. %as a counterpart. 
%Taking this into account, the NuSTAR observations are excluded from our %analysis because they exhibit a remarkable K$_\alpha$ Fe line \cite{2022Pike} which is centered on $6.4$ keV with no significant redshift.

%, and thus likely correspond to a source unrelated to the gamma-ray emission.

\begin{figure}
\centering
\includegraphics[angle=0,width=8cm, angle=0]{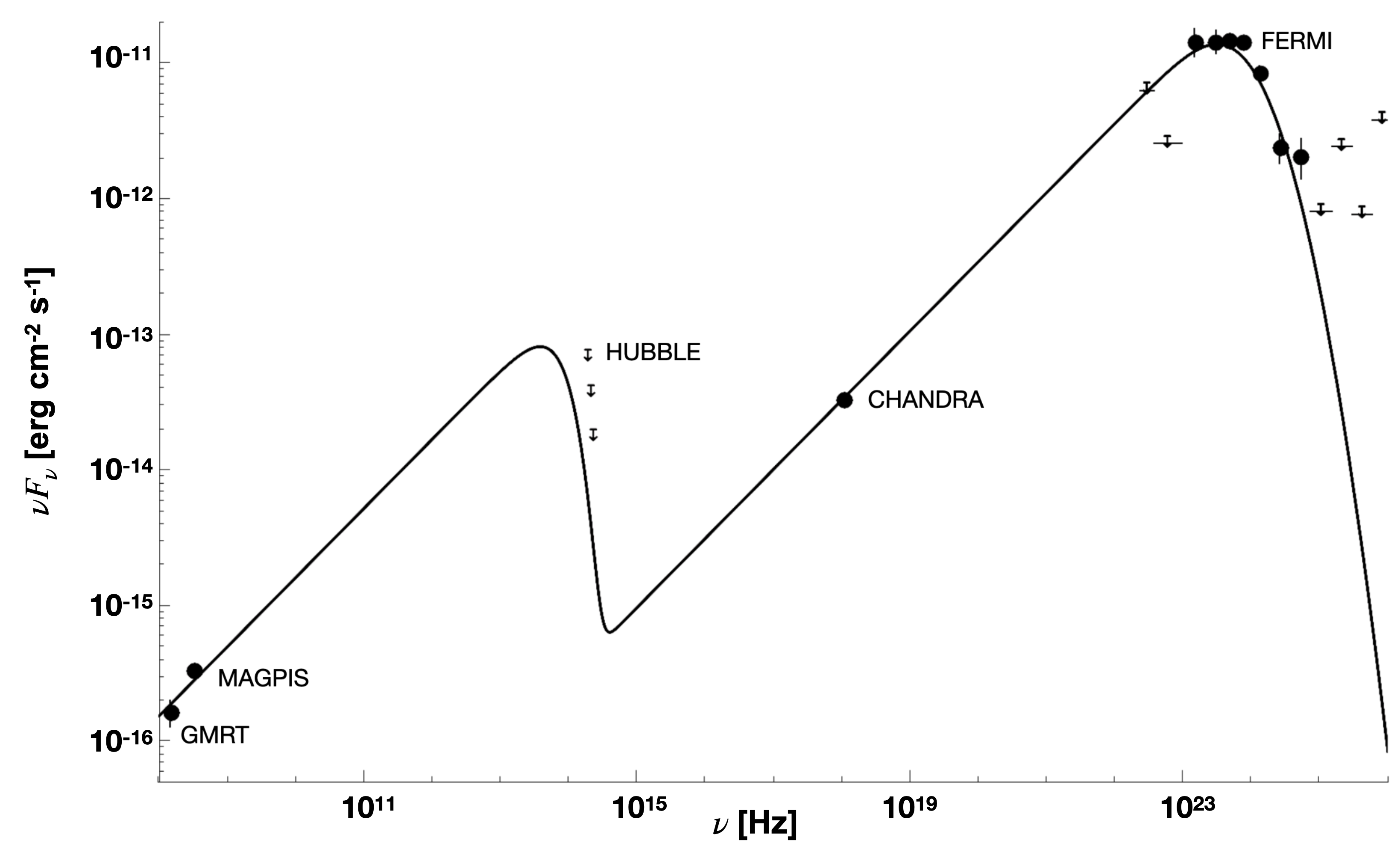}
\caption{Spectral Energy Distribution for the proposed blazar. Tentative fit with typical double bump is shown.}
\label{fig:SED}
\end{figure}

In an attempt to better characterize the agreement of the newly proposed blazar with its expected SED, we have assembled all the available multi-wavelength flux densities measurements in Fig. \ref{fig:SED}. We tried to fit the SED points based on the \citet{2017Paiano} approach to unveil blazars among multi-wavelength counterparts of Fermi unassociated sources, adapted for monochromatic flux as a function of frequency \citep[see][]{2020Marti}.
%as follows:
%\begin{equation}
%    \nu F_\nu = \sum_1^2 A_i \left(\frac{\nu}{\nu_i}\right)^{1-\alpha} \exp\left\{-\frac{1}{2\sigma_i^2}\left[\log\left(1+\frac{\nu}{\nu_i}\right)\right]^2\right\}
%\end{equation}
Despite the scarcity of data and its lack of simultaneity, we see here that the consistency with the typical two-bump SED appearance seems plausible for a FSRQs. The MAGPIS 20 cm point is not appearing in the graphic because it does not follow the SED trend, which reinforces our previously suspicions of variability. A common spectral index $\alpha = 0.50 \pm 0.01$ for both synchrotron and Compton bumps resulted from the fit.

The blazar FSRQ hypothesis may be tested if we try to apply a classification method based on broad-band effective spectral indices, defined as $\alpha_{12}=-\frac{\log(f_2)-\log(f_2)}{\log(\nu_2)-\log(\nu_2)}$, where $f_1$ and $f_2$ are flux densities at $\nu_1$ and $\nu_2$, respectively. Following the work of \citet{2021Ouyang}, three energy bands are considered (radio at 5 GHz, optical at 5100 \AA, and X-ray at 1 keV) to calculate two effective spectral indices ($\alpha_{ro} = 1.3$ and $\alpha_{ox}=0.9$), converting the observed data to those wavelengths by assuming power laws with canonical empirical effective spectral indices for radio, optical and X-rays \citep[as, for instance, in ][]{Abdo2010_SED}. As in our case the optical data is only a limit, so will be the corresponding $\alpha_{ro}$ and $\alpha_{ox}$. In spite of that, the classification of the source according to the criteria in \citet{2021Ouyang} is always a FSRQ.

% un ajuste con igual alfa sería alfa=0.52 (ten en cuenta que ya es negativo en el formalismo de Paiano) y s1 0.1 y s2 0.9 y A1 2 y A2 190 y nu2 399950000000000000000000 y nu1 4.714x
A further consistency check to our blazar hypothesis comes from energetic considerations. The observed gamma-ray flux in the Fermi band ($50$ MeV $- 1$ TeV) is $4.22 \times 10^{-8}$ photons cm$^{-2}$ s$^{-1}$ for \bicho. The modeled log-parabola SED of this source gives a reference energy of $E_{ref} = (2.91 \pm 0.04) \times 10^{3}$
MeV, so the gamma-ray flux may be expressed as $F_\gamma = (1.97 \pm 0.03) \times 10^{-10}$ erg cm$^{-2}$ s$^{-1}$. If we now assume a luminosity for \bicho\ of the order of the mean luminosity of bright Fermi blazars \citep[$1.38 \times 10^{47}$ erg s$^{-1}$, see][]{2010Ghisellini}, we obtain a luminosity distance of about $2420$ Mpc. Taking a Hubble constant of $H_0 = 69.6$ km s$^{-1}$ Mpc$^{-1}$, and assuming a flat Universe with $\Omega_{vac} = 0.714$, this luminosity distance is equivalent to a redshift $z \simeq 0.43$, which seems plausible \citep[\textit{e.g.}][]{2015arsioli, 2021Ouyang}. 
% AQUI USO ESTA PAGINA PARA CALCULAR https://www.astro.ucla.edu/~wright/CosmoCalc.html
At this distance the K-corrected X-ray luminosity, using a typical FSRQ spectral index $\alpha = -0.78$ \citep{2021Ouyang}, 
%OJO! Ten en cuenta que hemos definido alfa más arriba sin el signo, escribiendo nu^-alpha, así que un exponente positivo implica un alfa negativo
is $6.8 \times 10^{42}$ erg s$^{-1}$ for the narrow 2 \textendash 10 keV  range, while for radio, with a flat spectral index, we get $4.9 \times 10^{43}$ erg s$^{-1}$. These values seem to be compatible with those of blazars \citep[\textit{e.g.}][]{2020Costamante}. If, on the other hand, our source were an isolated MSP inside G01 at 5 kpc, its corresponding X-ray and radio luminosities were $3.1 \times 10^{31}$ erg s$^{-1}$ and $1.1 \times 10^{30}$ erg s$^{-1}$ \citep[assuming a typical radio spectral index of $\alpha = 1.4$, ][]{2004lorimer, 2013bates}, respectively. While the first one may be reasonable for an MSP \citep{Lee_2018}, the radio luminosity is a bit higher than the mean one for this kind of sources \citep{2014Szary}. This seems to be against the MSP nature of our X6 source, especially if we take into account that, as was pointed out before, the observed radio spectral index is flat instead of a steep value. Moreover, it is very difficult to explain the observed hints of long time-scale variability for an isolated, old MSP with no expected accretion winds. 
Therefore, all the observational evidences taken together tip the scale in favor of the extragalactic origin of our source.

\section{Conclusions}
Here we have presented a plausible blazar coincident with \bicho\ that may be contributing to its gamma-ray emission, thus proposing an alternative nature for this source. To reveal if this is the only high-energy emitter or if \bicho\ is the result of a collective contribution is out of the scope of this paper.
Future X-ray, gamma-ray and radio observations will hopefully be able to proof some sort of correlated variability that unambiguously associates all emissions with a common origin.
Assuming that our alternative identification is correct, the number of MSP in G01 as estimated by \citet{2022Wu}, or their particle acceleration efficiency, would need to be reduced.

%For our purposes, the two distributions compared are the measured distribution of arrival times, and the accumulated fraction on the time interval elapsed. If the null hypothesis (no variability) holds, we get 50% of your events in 50% of the passed time, and the statistic D (= error absoulto máximo entrae ambas distribuciones) should distribute according to the Kolmogorov distribution for many realizations of the arrival times distribution. The probability returned by the test is thus pKS=1−α, where α is the probability (under the Kolmogorov distribution) that the value of D is larger or equal than the measured value. A small value of pKS therefore indicates consistency with the null hypothesis, whereas a large value of pKS indicates that the interval between events is not constant, and therefore variability can be inferred.

\section*{Acknowledgements}

JAC and JFAC are CONICET researchers. JAC is a Mar\'ia Zambrano researcher fellow funded by the European Union -NextGenerationEU-  (UJAR02MZ). JAC and JFAC  acknowledge support by PIP 0113 (CONICET) and PICT-2017-2865 (ANPCyT). PLLE, JM, EM and JAC were also supported by grant PID2019-105510GB-C32/AEI/10.13039/501100011033 from the Agencia Estatal de Investigaci\'on of the Spanish Ministerio de Ciencia, Innovaci\'on y Universidades, and by Consejer\'ia de Econom\'ia, Innovaci\'on, Ciencia y Empleo of Junta de Andaluc\'ia as research group FQM- 322, as well as FEDER funds.

%%%%%%%%%%%%%%%%%%%%%%%%%%%%%%%%%%%%%%%%%%%%%%%%%%
\section*{Data Availability}

All data used in this work is public, and available in their respective databases.



\bibliographystyle{mnras}
\bibliography{misrefs} % if your bibtex file is called example.bib

%\begin{thebibliography}{99}
%\end{thebibliography}

%%%%%%%%%%%%%%%%%%%%%%%%%%%%%%%%%%%%%%%%%%%%%%%%%%

%%%%%%%%%%%%%%%%% APPENDICES %%%%%%%%%%%%%%%%%%%%%

%\appendix



%%%%%%%%%%%%%%%%%%%%%%%%%%%%%%%%%%%%%%%%%%%%%%%%%%

% Don't change these lines
\bsp	% typesetting comment
\label{lastpage}
\end{document}